\documentclass[11pt]{article}
\usepackage[margin=1.2in]{geometry}
\usepackage{amsfonts}
\usepackage{fancyhdr}
\usepackage{amsmath}
\usepackage{amssymb}
\usepackage{amsthm}
\usepackage{setspace}
\usepackage{hyperref}
\usepackage{enumitem}
\usepackage{natbib}
\usepackage{comment}
\usepackage{gensymb}
\usepackage{url}

\newcommand{\counterfactual}{%
    \mathrel{\mathop\Box}\mathrel{\mkern-2.5mu}\rightarrow
}

\usepackage{xcolor}

\title{Without microphysical causation, just anything cannot begin to exist just anywhere}

\author{Daniel Linford}

\begin{document}

\maketitle

\begin{abstract}
    \noindent According to the Causal Principle, anything that begins to exist has a cause. In turn, various authors -- including Thomas Hobbes, Jonathan Edwards, and Arthur Prior -- have defended the thesis that, had the Causal Principle been false, there would be no good explanation for why entities do not begin at arbitrary times, in arbitrary spatial locations, in arbitrary number, or of arbitrary kind. I call this the Hobbes-Edwards-Prior Principle (HEPP). However, according to a view popular among both philosophers of physics and naturalistic metaphysicians -- Neo-Russellianism -- causation is absent from fundamental physics. I argue that objections based on the HEPP should have no dialectical force for Neo-Russellians. While Neo-Russellians maintain that there is no causation in fundamental physics, they also have good reason to reject the HEPP.
\end{abstract}

\noindent \textbf{Penultimate draft.} Please cite published version.

\section{Introduction}

Bertrand Russell (\citeyear{Russell:1912}) famously maintained that mature science is incompatible with causation. Not only do mature sciences fail to mention causation, but the dependence relations that do appear in mature science are bad candidates for interpretation as causal relations. For that reason, Russell advises us to eliminate causation altogether. While the majority of philosophers no longer find Russell's eliminativism compelling, many contemporary naturalistic metaphysicians and philosophers of physics are \emph{Neo-Russellians}. Neo-Russellianism is the conjunction of two theses: (i) causation is not physically fundamental and (ii) there are real (but not fundamental) causal relations that help guide effective strategies \citep{Blanchard:2016}. For Neo-Russellians, causal relations can (for example) help guide successful medical interventions, even though all of the causal relations are underwritten by more fundamental (and so acausal) goings-on.

According to the \emph{Causal Principle}, anything that begins to exist has a cause. A number of authors -- including Thomas Hobbes (\citeyear[39-40]{Hobbes:1999}), Jonathan Edwards (\citeyear{Edwards:1962}), and Arthur Prior (\citeyear{Prior:1962}) -- have argued that if the Causal Principle is false, there is no relevant difference between the kind or number of entities or their spatial or temporal location that explains why only \emph{those} entities begin to exist and not others. Hence:

\begin{quote}
    \textbf{The Hobbes-Edwards-Prior Principle (HEPP):} If the Causal Principle is false, then there would be no good explanation for why entities do not begin at arbitrary times, in arbitrary spatial locations, in arbitrary number, or of arbitrary kind.
\end{quote}

\noindent If entities can begin without being caused, the thinking is supposed to go, just any kind and any number of things could just begin to exist at just any time and in just any place and we'd have no good explanation for why they do not. In other words, if things can begin without causes, nothing at all could select the times, spatial locations, number, or kind of things that begin to exist; moreover, nothing at all could prevent objects from appearing at arbitrary times, in arbitrary spatial locations, in arbitrary number, or of arbitrary kind. Our world doesn't exhibit such arbitrariness; hence, the observed order in our world strongly supports the Causal Principle.

The HEPP is utilized as part of a reductio for those who -- like the Neo-Russellians -- deny the Causal Principle. My aim in this essay is to show a reductio of that kind has no dialectical force for Neo-Russellians. My aim is not to offer a general defense of Neo-Russellianism, which has already been done at length elsewhere. In section \ref{the-argument}, I lay out my general argumentative strategy. Afterwards, in section \ref{kinds-of-NR}, I briefly summarize four routes to Neo-Russellianism, each of which can be associated with its own version of Neo-Russellianism. In section \ref{causal-principle-probably-false}, I show that if Neo-Russellianism is true, then there is good reason to think the Causal Principle is false. This suffices for establishing that there is at least tension (if not outright contradiction) between Neo-Russellianism and the Causal Principle. In section \ref{hobbes-edwards-prior}, I briefly summarize the case for the HEPP as offered by Hobbes, Edwards, and Prior. Lastly, in section \ref{non-causal-explanations}, I will show that there are adequate non-causal resources Neo-Russellians can utilize to explain why entities do not begin at arbitrary times, in arbitrary spatial locations, in arbitrary number, or of arbitrary kind. Hence, anyone who has good reason to accept Neo-Russellianism has good reason to reject the HEPP.

\section{The Argument\label{the-argument}}

According to the Causal Principle, anything that begins to exist has a cause. Let's denote the Causal Principle as $C$. There are at least three different interpretations of $C$. First, the claim might be that, as a matter of logical necessity, anything that begins to exist has a cause. This interpretation is presumably too strong. For example, the negation of $C$ does not appear to result in a contradiction. Second, the claim might merely be that, at the actual world, everything that begins has a cause, even though this isn't true of logical, metaphysical, or physical necessity. On this interpretation, $C$ is an accidental generalization. Not only does $C$ appear too weak on this interpretation, but there might be no way to establish $C$ either through an inference to the best explanation, inductive argumentation, or logical demonstration. Lastly, we can interpret $C$ as the claim that, as a matter of metaphysical or physical necessity, anything that begins to exist has a cause. I think this interpretation is the most plausible and so I will assume this interpretation going forward.

Let's use $F$ to represent the sentence `entities do not begin at arbitrary times, in arbitrary locations, in arbitrary number, or of arbitrary kind'. Moreover, let's use $E(\cdot)$ to represent the monadic predicate `there is a good explanation of'. The HEPP states that if $C$ is false, then there would be no good explanation for $F$. The HEPP is either a would-counterfactual conditional of the form $\neg C \counterfactual \neg E(F)$ or else a would-counterpossible conditional. For that reason, the HEPP is false just in case the closest worlds where the antecedent is true -- that is, the closest worlds where $\neg C$ is true -- are worlds where the consequent is false -- that is, worlds where $\neg E(F)$ is false.

The Neo-Russellian should think that $C$ is false or only vacuously true. Either way, the Neo-Russellian has good reason to reject the HEPP. First, there's the possibility that Neo-Russellians should think $\neg C$ is true. As I discuss in section \ref{causal-principle-probably-false}, the Neo-Russellian might have good reason for thinking that $\neg C$ is true. However, as discussed in section \ref{non-causal-explanations}, the Neo-Russellian should think there is a good explanation for $F$ in the actual world. If $\neg C$ is true and $\neg E(F)$ is false, then the HEPP is false. Now consider the possibility -- also discussed in section \ref{causal-principle-probably-false} -- that Neo-Russellians should think $C$ is vacuously true. If $C$ is vacuously true, then the HEPP is false just in case, at the closest worlds where $\neg C$ is true, $\neg E(F)$ is false. Depending upon whether the HEPP is a counterfactual conditional or a counterpossible conditional, those worlds will either be possible or impossible. Regardless, again, section \ref{non-causal-explanations} shows that important non-causal explanations are available in the absence of $C$, and hence would be available in the closest worlds where $C$ is false. Hence, the Neo-Russellian should conclude that the HEPP is false. Since the Neo-Russellian should conclude that the HEPP is false, objections based on the HEPP do not have dialectical force for the Neo-Russellian.

\section{\label{kinds-of-NR}Four Routes to Neo-Russellianism}

In this section, I briefly summarize each of four routes to Neo-Russellianism. Each route to Neo-Russellianism results in a variant of Neo-Russellianism; some of the variants (perhaps not all) can be combined, so the variants are not necessarily exclusive. Let's begin by putting the microphysical/macrophysical distinction on the table. Suppose that $O$ is some physical system with a large number of constituent parts; $O$ might be a desk, a chair, or even a cat. The microphysical description of $O$ specifies $O$'s exact state, such as the positions and momenta of all the particles in a cloud of Newtonian particles. In contrast, macrophysical descriptions result from coarse-graining over the microphysical details; for the cloud of Newtonian particles, a macrophysical description may include the cloud's pressure, volume, mass, and the approximate spatial region occupied by the cloud. The four routes to Neo-Russelllianism have the following structure:\footnote{There are some additional arguments for Neo-Russellianism that do not fit this structure. For example, David Papineau (\citeyear{Papineau:2013}) and Mark Pexton (\citeyear{Pexton:2017}) have argued that Neo-Russellianism and related views provide a simple solution to the Generalized Causal Exclusion Problem \citep[e.g.,][]{Block:2007}.}

\begin{enumerate}
    \item The microphysics has feature $F$.
    \item If causation were included in the microphysics then the microphysics would not have feature $F$.
    \item Therefore, causation is not included in the microphysics.
\end{enumerate}

\noindent The four routes to Neo-Russellianism fill out $F$ in distinct ways, corresponding to four variants of Neo-Russellianism. The general theme is that whatever microphysical dependency relations there are so radically violate our causal intuitions that continuing to refer to such relations as \emph{causal} relations would be misleading and inappropriate.

\textbf{Neo-Russellianism-1}. The past nomologically determines the future in the same way the future determines the past \citep{Russell:1912, Field:2003, albert:2003, FarrReutlinger:2013, Loewer:2007b, Loewer:2012b, loewer:2020, Adlam:2023, Rovelli:2023, Ismael:2023}. We intuitively expect causation to be asymmetric, with effects depending on causes and causes preceding effects. If past and future microphysically determine each other symmetrically, there is no microphysical distinction between causes and effects. Without a distinction between causes and effects, there is no microphysical causation.

One could object that causal relations can be symmetric. To cite a stereotypical example, if one book rests on another, so that neither book falls, then both books are simultaneous causes of each other's failing to fall. Set aside the fact that, in relativistic physics, no events in spatiotemporally disjoint regions can be objectively simultaneous. Even if two events (or whatever) can be the simultaneous causes of each other, such occurrences are understood to be rare. Moreover, not only are the determination relations found in fundamental physical theory usually symmetric, but they are never asymmetric. So, the problem is not that what is common in fundamental physics is uncommon for causation or vice versa; the problem is that one ubiquitous feature of causation -- asymmetry -- never shows up in fundamental physics.

\textbf{Neo-Russellianism-2}. Paradigmatically, causes and their effects are local events; as examples, Thomas Blanchard (\citeyear[257]{Blanchard:2016}) discusses the throwing of a rock to break a window and the lighting of a match to cause a forest fire. We can add that, in paradigmatic cases, some past events contribute more so than do others, e.g., events in the American Senate in 1855 contributed more to the American Civil War than did events in Titan's orbit in 1700. In contrast, microphysically, each event equally counterfactually depends upon all of the events in some large region (\citeauthor{Blanchard:2016}, \citeyear[257]{Blanchard:2016}; \citeauthor{Latham:1987}, \citeyear{Latham:1987}). For example, in relativistic spacetimes, events are determined by their entire past light cone or, in the case of a globally hyperbolic spacetime, all of the data on a past or future Cauchy surface. Laws can also be understood to relate entire regions one to another, e.g., one Cauchy surface might be determined by another Cauchy surface. Blanchard (\citeyear[257]{Blanchard:2016}) concludes that ``there is a mismatch between the sort of states involved in nomic relations and the kinds of entities that we regard as prototypical causes''.

Perhaps one could object as follows. If an event equally microphysically depends on all the events to its past, then, instead of denying microphysical causation, we should think that the total cause of an event is just all of the prior events. To help see why friends of Neo-Russellianism-2 would encourage us not to make that move, consider a variation on a reply Alfred Mele (\citeyear[253]{mele:1992}) made in response to non-causal accounts of libertarian free-will. Suppose that an agent's intention to $\phi$ supervenes on a specific neural circuit. In that case, a mad scientist might arrange for one of agent $A$'s intentions to $\phi$ to supervene on neural circuit $N_1$, another one of $A$'s intentions to $\phi$ supervene on neural circuit $N_2$, and might also arrange for $N_1$ to be causally screened off with respect to $A$'s $\phi$ing. In that case, there's a strong intuition that, though $A$ has both intentions, only $N_2$ explains why the agent $\phi$ed; furthermore, $N_1$ and $N_2$ have these distinct roles precisely because $N_2$, and not $N_1$, caused $A$ to $\phi$.

But, microphysically, all events to the past of $A$'s $\phi$ing equally contribute to $A$'s $\phi$ing. For that reason, microphysically, it's not available for us to say that $A$'s $\phi$ing microphysically depends upon $N_2$ in a way that $A$'s $\phi$ing did not depend upon $N_1$. Perhaps worse, we can imagine that $A$ has a third neural circuit $N_3$ and that an intention not to $\phi$ supervenes on $N_3$. Surely, $A$'s $\phi$ing should not be explained by a neural circuit realizing $A$'s intention not to $\phi$. Yet, given the microphysical laws, $N_1$, $N_2$, and $N_3$ all have equal priority in explaining $A$'s $\phi$ing. 

Re-describing the microphysics in causal terms can only ever mislead; insofar as intentions cause actions, we certainly should think that $N_1$, and not $N_2$ or $N_3$, causes the agent to $\phi$. But that statement is available only after we've appropriately coarse grained over microphysical degrees of freedom and so only at the macrophysical level. Hence, appropriate causal ascriptions are available only at the macrophysical level.

Unlike the previous two, the next two routes are not direct descendants of Russell's original arguments. However, I include them here because, as with more traditional forms of Neo-Russellianism, they deny that causation is fundamental or irreducible.

\textbf{Neo-Russellianism-3}. Some philosophers have considered nomological indeterminism a reason to deny microphysical causation. For example, Friedrich Waismann (\citeyear{Waismann:1961}) and Bradley Monton (\citeyear{Monton:2009}) have argued that indeterminism in quantum mechanics provides a reason to deny microphysical causation, while John Norton (\citeyear{Norton:2003}) delivers an argument based on indeterminism in classical mechanics. The equations fundamental to General Relativity -- the Einstein Field Equations -- likewise include indeterministic solutions \citep[see, for example, ][]{Earman:1995} that could be leveraged for an argument against microphysical causation. 

One could point out that philosophers now recognize indeterministic causation. In that case, it may be that physical indeterminism is not a good reason to give up on microphysical causation. However, at least with respect to free-will, philosophers also recognize that there are two categories of indeterminism; for example, an agent's intentions could be either indeterministically caused or could be indeterministically uncaused \citep{Ginet:2002}. Likewise, there's room enough for two interpretations of physical indeterminism, one where indeterministic events are uncaused and another where they are indeterministically caused. Consequently, unless we have a good argument for favoring the former interpretation over the latter, both are live options. As far as I know, a convincing argument of that sort has not been forthcoming. Since a convincing argument of that sort has not been forthcoming, Neo-Russellianism-3 remains a live option.

\textbf{Neo-Russellianism-4}. Thomas Kuhn (\citeyear{Kuhn:1977}) and Michael Redhead (\citeyear{Redhead:1990}) argued that all motion is natural motion. Aristotle understood natural motion as the motion inanimate objects undergo absent efficient causes. Early modern physicists viewed inertial motion similarly, with only accelerated motion having an efficient cause. For Kuhn and Redhead, contemporary physics has completely eroded the distinction between natural and non-natural motion. Redhead argues that this is particularly apparent in General Relativity, where motion that was once attributed to a force is now attributed to inertial motion in a curved spacetime. Kuhn and Redhead's reasons for denying any motion has a cause are generalizable. According to the Aristote-inspired view endorsed by Redhead and Kuhn, objects undergo natural motion – such as inertial motion – in virtue of what those objects intrinsically are and in the absence of efficient causes. Likewise, perhaps, absent efficient causes, objects undergo natural changes in virtue of what those objects intrinsically are. In turn, perhaps explanations in fundamental physics just are explanations of the natural changes that fundamental entities undergo. In that case, we are left with an elimination of causation from fundamental physics.\footnote{Related points have been made in Menzies (\citeyear{Menzies:2007}) and Bunge (\citeyear[112-114]{Bunge:1979}).}

The four routes to Neo-Russellianism are not necessarily compatible and philosophers may defend one (or more) without endorsing the others. For example, Neo-Russellianism is often understand as the combination of Neo-Russellianism-1 and -2. Some indeterministic physical theories, e.g., objective collapse theories, conflict with the view that nomological determination is temporally symmetric. Furthermore, each route faces objections. Quantum mechanics may suggest indeterminism, but deterministic interpretations like the many-worlds interpretation or Bohmian mechanics exist.  The collapse of the distinction between natural and non-natural motion remains controversial; for example, Einstein’s goal to unify natural and non-natural motions in General Relativity is widely seen as unfulfilled \citep{Hoefer:1994}. Still other philosophers will reject all four routes by responding that the motivations for each of the four result only from our causal intuitions misfiring. Perhaps we need to develop new causal notions. For example, perhaps -- as John McTaggart (\citeyear[224-227]{McTaggart:1921}; \citeyear[182]{McTaggart:1927}) held -- we should revise our conception of causation by removing the asymmetry between causes and effects. For the purpose of this article, I am setting aside a complete defense of Neo-Russellianism. This paper's goal is more modest: to show that objections to Neo-Russellianism based on the HEPP are ineffective.

\textbf{A further clarification.} Russell maintained that we should eliminate causation altogether. However, philosophers defending Neo-Russellianism do not usually deny causation entirely. We need causation to explain phenomena in the special sciences, and, as Nancy Cartwright (\citeyear{Cartwright:1979}) famously and convincingly argued, we need causation to determine effective strategies. As with many phenomena that appear at the macrophysical but not the microphysical level, causation could be reducible to, emergent from, or otherwise supervenient on the microphysics. Alternatively, causation could be perspectival — a real pattern that becomes salient for agents embedded within the physical world, though not a fundamental feature of physical reality.

\section{\label{causal-principle-probably-false}Neo-Russellianism and the Causal Principle}

In this section, I will show that, provided any one of the four versions of Neo-Russellianism, we have good reason to reject $C$, that is, the principle that anything that begins to exist has a cause. To show that any one of the four conflicts with $C$, it's enough to show that, given any one of the four, there's a physically possible world where an entity begins without a cause. At first glance, this may seem easy enough. Neo-Russellianism tells us that, microphysically, there are no genuine causal relations. Provided that at least one microphysical entity begins to exist, $C$ is false; some entities begin to exist without causes.

However, some readers might be inclined to think that if Neo-Russellianism is true, then $C$ will turn out to be trivially true. Many Neo-Russellians think that there is no microphysical direction of time. Whether any item begins to exist should depend upon whether that item has finite extension in the earlier-than direction. If there is no microphysical direction of time, then no microphysical entities have a finite extension in the earlier than direction; hence, no microphysical entities begin to exist. In that case, $C$ is trivially satisfied microphysically. Macrophysically, there is a direction of time, but there are also causes -- so, $C$ is macrophysically satisfied as well. Call this the \emph{Trivialism Reply}.

As an anonymous reviewer has suggested, some readers will be completely undaunted. It's enough, one might think, that there is a direction the macrophysical arrow of time identifies as the past and that various microphysical entities have finite temporal extension in that same direction. In that case, since there is no microphysical causation, such entities begin without causes. However, other readers -- who may have a more substantive conception of beginning to exist -- will find this unconvincing.

Neo-Russellians ought to point out that, if their view includes a reductive account of the direction of time, then it's nomologically possible for a macrophysical entity to begin to exist without a cause. For example, consider Ludwig Boltzmann's cosmological view in which the whole of physical reality is a gas at thermodynamic equilibrium. For Boltzmann, there are local departures from thermodynamic equilibrium -- places where the entropy fluctuates to some small value -- and a locally defined arrow of time, pointing away from the entropy minimum. We can then use that locally defined arrow of time to locally define a corresponding B series. The corresponding macrophysical B series begins at the entropy minimum, with no possibility for anything temporally prior.

If anything begins at a local entropy minimum -- a Boltzmann Brain, for example -- then whatever begins at the entropy minimum couldn't have a cause prior to the entropy minimum simply because, with respect to the locally defined B series, there is nothing before. Instances where there is a macrophysical direction of time, where things begin to exist, but also where there is no macrophysical causation are nomologically possible. Hence, the Trivialism Reply fails.

Neo-Russellianism-2, -3, and -4 are neutral with respect to whether there is a microphysical direction of time. Hence, Neo-Russellians can in principle avoid the Trivialism Reply by adopting a microphysical direction of time. However, even if they do not adopt a microphysical direction of time, Neo-Russellianism-3 and -4 have additional resources for rejecting the Trivialism Reply. For example, consider Neo-Russellianism-3, in which the microphysics is indeterministic. Though the occurrence may be exceedingly rare, if a sufficiently large swarm of microphysical entities indeterministically begins, there may be a corresponding macrophysical object that begins without any causal antecedents. Neo-Russellianism-4 offers resources for rejecting the Trivialism Reply as well. Fundamental particles are discrete excitations in fundamental fields, e.g., an electron is an excitation of the electron field. In turn, compound objects supervene on complicated field configurations. If the changes that fundamental entities undergo are perfectly natural for those entities -- as they are on Neo-Russellianism-4 -- then fundamental fields only undergo changes that are perfectly natural for them. Hence, microphysically, when a new object comes into being -- i.e., when fundamental fields evolve into some new configuration -- no microphysical causation is involved.

\section{\label{hobbes-edwards-prior}Hobbes, Edwards, and Prior}

So far, I've shown that if Neo-Russellianism is true, then $C$, the principle that anything that begins to exist has a cause, is either false or trivially true. However, even if $C$ is trivially true, the closest worlds where $C$ is false are worlds that satisfy the actual laws. As we'll see in section \ref{non-causal-explanations}, this is crucially important, because it means that in the nearest worlds where $C$ is non-trivially false, there are still adequate explanations for what does, and what does not, begin to exist. In this section, I turn to considering how the HEPP may be used in arguments defending $C$; subsequently, I'll show that these arguments are not convincing to the Neo-Russellian because, in the closest worlds where $\neg C$ is true, the HEPP is false. 

Friends of the HEPP maintain that good explanations for the times, locations, kinds, or number with which things begin to exist require $C$ because, without $C$, all times, locations, numbers, and kinds would be on a par, with nothing to prefer any one in particular. For example, Hobbes argues that there would be no sufficient reason for something to begin at one time as at any other. According to Hobbes, if we try to imagine an item beginning without a cause, we find

\begin{quote}
    as much reason, if there be no cause of the thing, to conceive it should begin at one time as at another, that is, he has equal reason to think it should begin at all times; which is impossible, and therefore he must think there was some special cause why it began then rather than sooner or later; or else that it began never, but was eternal \cite[39-40]{Hobbes:1999}.
\end{quote}

\noindent Hobbes thought it was impossible for an item to have equal reason to begin at one time as at any other because, in cases where an event can happen in more than one way, the event will occur in all possible ways. For example, when considering a thing beginning to move without being caused to begin moving, Hobbes argues that there would be equal reason for the thing to begin moving in every direction. In that case, the thing would move in all possible directions at once -- which is impossible \cite[115]{Hobbes:1962}. Similarly, if a thing can stop moving without a cause, then -- once more -- nothing would favor one instant over any other, so that ``its motion would cease in every particle of time alike'' \cite[116]{Hobbes:1962}. Likewise, if something could begin to exist without a cause, there would be equal reason for that thing to begin to exist in any instant. Since a thing cannot possibly begin to exist in more than one instant, Hobbes concluded that a thing cannot begin to exist without a cause. When responding to Hobbes, Hume (\citeyear[56-57]{Hume:2005}) added that, without $C$, items could begin at just any spatial location. This a clear and obvious extension of Hobbesian logic: without $C$, there would be equal reason for an item to begin at any one spatial location as at any other.

For contemporary readers, Hobbes's impossibility claim is bizarre. For example, in David Lewis's combinatorial conception of modality, that $x$ can happen in multiple ways means only that $x$ happens in a distinct way in multiple worlds; Lewis would deny that there is one world where $x$ happens in multiple ways. In any case, Hobbes clearly leads us to a portion of the HEPP: without $C$, there would be no explanation for why an entity would begin at any one time or place as opposed to any other.

Jonathan Edwards added that, absent $C$, nothing could select among the many possible kinds of entities that might begin to exist: “stones or stars, or beasts, or angels, or human bodies, or souls, or only some new motion or figure in natural bodies, or some new sensations in animals, or new ideas in the understanding, or new volitions in the will; or anything else of all the infinite number of possibles” \cite[184]{Edwards:1962}. Why did Edwards maintain that, without $C$, nothing would select the kinds of things that begin to exist? One way to explain the existence of an item involves the nature of the item, itself. But -- Edwards thinks -- the item's nature cannot explain why only items of its kind began to exist. Before the item begins to exist, its nature does not yet exist -- and once it exists, it is too late for its nature to explain its coming into being. Hence, whatever selects something for existence must exist prior to it. However, if something begins to exist without a cause, nothing prior selects it. This is not to reify nothing -- the claim is not that there is an entity, \emph{nothing}, that selects what kinds of entities begin. Instead, there are \emph{no entities at all} that determine that one possibility is realized as opposed to another. As Edwards puts it, ``contingence is blind and does not pick and choose for a particular sort of events'' \cite[184]{Edwards:1962}. He concludes that, without $C$, entities would begin to exist wherever they are not prevented from doing so. Given the vast number of times and places where this is possible, an arbitrary and potentially limitless number of entities could -- or even would -- spontaneously come into existence.

Prior adopts a version of the Hobbes/Edwards view that, without $C$, there would be no reason why only some items begin to exist and not others. Unlike Hobbes, for Prior, the issue concerns explanation. For example, he tells us that it would be ``incredible that they should all turn out to be objects of the same sort'' \cite[61]{Prior:1962}. I take it that the resulting state of affairs would be merely ``incredible'', as opposed to incoherent, because, while wildly implausible, the resulting state of affairs does not involve a contradiction.

One may worry that there is only a verbal dispute between Hobbes, Edwards, Prior, on the one hand, and Neo-Russellians on the other. Neo-Russellians can consistently maintain that, though there is no microphysical causation, there are sufficient reasons for all microphysical events. Importantly, Edwards tells us that he is using the term `cause' more generally than the ordinary sense, where the ordinary sense refers to something with ``positive efficiency or influence to produce a thing, or to bring it to pass'' \cite[180]{Edwards:1962}. In fact, for Edwards, a cause is merely the ``nature of a ground or reason why some things are, rather than others; or why they are as they are, rather than otherwise.'' If, by `causation', Hobbes, Edwards, and Prior only mean sufficient reasons, then perhaps their disagreement is merely verbal.

There is more than a verbal difference between the two sides. First, while Edwards argues at length that causation is inconsistent with indeterminism \cite[213-216]{Edwards:1962}, Neo-Russellianism-3 entails indeterminism. Second, Neo-Russellianism-1 entails that the sufficient reason for an event to take place might not be temporally prior to the event. In contrast, Edwards makes clear that, in his sense of `cause', causes are antecedent to their effects -- either in the order of time or in the order of nature \cite[178, 191]{Edwards:1962}. Edwardsian causes are asymmetric, with effects that \emph{depend on} their causes \cite[183, 185]{Edwards:1962} and causes that necessitate their effects by being past or simultaneous with them \cite[153-154]{Edwards:1962}. Furthermore, Edwards's (\citeyear[171-174]{Edwards:1962}) central argument against libertarian free-will presupposes that causation is transitive and asymmetric.

Third, Edwards's conception of causation is similar to Hume's conception: for $A$ to cause $B$ involves a constant conjunction of $A$ and $B$.\footnote{Edwards adds that $A$ determines and necessitates $B$ \cite[153-154]{Edwards:1962}. God ordains and so necessitates the constant conjunctions \cite[124-127]{Harris:2005}.} For Neo-Russellianism-2, the laws of nature involve precisely specified, large portions of physical reality, such as an entire Cauchy surface. As Russell pointed out, this is disastrous for any view in which causation involves constant conjunction. The notion that there are constant conjunctions between causes and their effects requires vagueness in both. Science begins with vague statements, e.g., that bodies fall, but progresses with precisified statements, e.g., the equation of motion of a falling body. Nonetheless, ``[a]s soon as the antecedents have been given sufficiently fully to enable the consequent to be calculated with some exactitude, the antecedents have become so complicated that it is very unlikely they will ever recur'' \cite[8-9]{Russell:1912}. If a precisely specified antecedent is unlikely to recur, then precisely specified antecedents are unlikely to appear in constant conjunctions. Hence, Edwardsian causation is distinct from the non-causal determination relations Neo-Russellianism-2 endorses.

Edwards and Prior left a few remarks suggesting a way Neo-Russellians can reply to objections based on the HEPP. As Edwards and Prior admit, without $C$, there would still be constraints on what can begin to exist. For example, Edwards (\citeyear[184]{Edwards:1962}) maintains that, without $C$,  entities of arbitrary kind would begin to exist ``where there is room for them, or a subject capable of them, and that constantly, whenever there is occasion for them.'' Edwards's comments implicitly suggest that where there is no room for them, or a subject capable of them, or no occasion for them, items would not begin to exist. That is, even without $C$, there can be non-causal constraints on what begins to exist.

Prior constrains what can begin to exist without $C$ by restricting the scope of the HEPP. For Edwards, any contingent item must have begun to exist and so must have been caused; this principle applies equally to ``substances or modes, or things and the manner and circumstances of things''. For example, if a body is at rest, and spontaneously begins to move, there must be ``some cause or reason of this new mode of existence''. Contra Edwards, Prior argues that to say that a mode (such as a headache) can begin to exist is ``metaphysically obfuscating'' \cite[58] {Prior:1962}. For Prior, the nature a thing intrinsically possesses constrains the changes possible for it. Supposing that an electron can change its orbit without being caused to do so, the electron may only be able to jump to orbits $A$ or $B$ and not to some third orbit $C$. This depends not on the nature of the ``not-yet-existent jumps'' but rather ``the nature of the \emph{existing} electrons (and of other existing things)'' \cite[59]{Prior:1962}. Prior continues:

\begin{quote}
    It is rather that there are certain already-existing objects which have certain capacities, and some of which lack them, and none which have certain other capacities. Persons, say, have the power, without the necessity, of doing $X$ in certain circumstances; for oysters, on the other hand, doing $X$ may be necessary or impossible; and $Y$, say turning into a dragon, may be something which no existing object has the power to do. [...] it is \emph{their} limitations -- the limitations of actually present things, not those of still absent events -- which, while leaving some alternatives possible, do not leave all alternatives possible \cite[59-60]{Prior:1962}.
\end{quote}

\noindent Hence, for Prior, a metaphysics of substances with capacities and dispositions allows for a ``limited indeterminism'' without letting in all of the goblins Edwards feared. Prior goes on to consider the relationship between his view and some of the cutting-edge physical theorizing of his day. According to the now rejected steady state cosmology, hydrogen atoms are spontaneously created in the space between galaxies, so that the average density of the universe remains constant over cosmic time. Prior considers, and rejects, the possibility that there is no cause for the creation of the hydrogen atoms.

To reiterate, in Prior's view, indeterminism is limited by the capacities substances intrinsically possess. If hydrogen atoms are substances and begin causelessly, then substances themselves could begin without causes; in that case, nothing could prevent other substances from beginning without causes. Nonetheless, in the last sentence of his article \cite[61]{Prior:1962}, Prior leaves open an important possibility -- that hydrogen atoms are not substances with capacities. Indeed, we now know that hydrogen atoms supervene on a complicated configuration of various fundamental fields. Those fields might be understood as substances with their own capacities, so that, without $C$, they can develop a complex pattern of modifications -- a hydrogen atom -- that spontaneously begins to exist in some circumstance, even though other entities -- corresponding to other configurations of fundamental fields -- cannot begin to exist in the same circumstanaces. While I do not take a stand on whether any substantial metaphysic is correct, views with similar consequences to Prior's can be adopted. In the next section, I will show that, sans $C$, physical reality can be non-causally constrained in ways that either select various items for existence or prevents other items from existing.

\section{Non-Causal Explanations for Beginning to Exist\label{non-causal-explanations}}

In this section, I show that even if $C$ is false, there would still be a good explanation for $F$. Let's recall why this is important for my central argument. Since the HEPP is either a would-counterfactual condition or else a counterpossible conditional, the HEPP is false just in case, in the closest worlds where $\neg C$ is true, that is, where the Causal Principle is false, $\neg E(F)$ is false, that is, there is a good explanation for $F$. In turn, there are two cases to consider. In the first, $\neg C$ is actually true and there actually is a good explanation for $F$. In the second case, $\neg C$ is not actually true, but, in the nearest-by worlds where $\neg C$ is true, there is a good explanation for $F$. Regardless, what needs to be shown is that, absent $C$, there would be a good explanation for $F$. That's the task I turn to in this section.

There are two sets of facts that need to be explained to explain $F$. One set, that I call the \emph{Preventive Facts}, satisfies the schema `items do not begin to exist that are $\phi$', where $\phi$ can be filled in by `of arbitrary kind', `in arbitrary number', `at arbitrary spatial locations', or `at arbitrary temporal locations'. The other set, that I call the \emph{Selective Facts}, satisfy the schema `items begin to exist only of select $\psi$', where $\psi$ can be filled correspondingly.\footnote{Since the Selective Facts imply Preventive Facts, we may be able to get by with only one set of facts. For example, if items can only begin of select kind, then items cannot begin of arbitrary kind. Nonetheless, while the Selective Facts imply Preventive Facts, the Preventive Facts do not imply Selective Facts. There could be a true Preventive Fact without a corresponding Selective Fact.} For example, one Selective Fact states that items begin to exist only of select kinds. In other words, to explain the Preventive Facts, one must articulate why items do not begin to exist with some specific range of features, while, to explain the Selective Facts, one must articulate why the items that do begin to exist have some specific range of features. Proponents of the HEPP imagine that, had $C$ been false, nothing could preclude vast numbers of entities from beginning to exist for no reason at all. For that reason, proponents of the HEPP imagine that nothing other than causes could explain the Preventive Facts.

Though they reject $C$, Neo-Russellians can commit to general principles with broad scope that they can use to explain $F$. To see this, consider Neo-Russellianism-1, where the future is just as determined by the past as the past is by the future, or Neo-Russellianism-2, where future (or past) events are determined by a large spacetime region. The universe can be determined, in either sense, even if $C$ is false. However, friends of Neo-Russellianism-3 insist that the entire history of the universe is not determined. Instead, events might merely be (for example) probabilified. To capture as many versions of Neo-Russellianism as possible, consider a schematic principle whose details can be filled in with whatever other commitments Neo-Russellians may have:

\begin{quote}
    \textbf{The Consistency Principle (TCP):} Whatever begins to exist must be consistent with the full set of logical, metaphysical, and nomic principles.
\end{quote}

\noindent The full set of logical principles are all of the inviolable logical truths. For example, an entity cannot begin to exist if, together with whatever already exists, a contradiction would result from its beginning to exist. The full set of metaphysical principles are all the metaphysical necessities there are. Lastly, the full set of nomic principles are all of the fundamental physical laws, whatever they turn out to be. Because they commit to the view that TCP is non-trivially satisfied, the Neo-Russellian claims to have a good non-causal explanation for $F$; the Preventive and Selective Facts can be explained non-causally by either logical principles, metaphysical principles, or by physical law.

Moreover, the TCP allows for both Neo-Humean and Anti-Humean accounts of laws. Taking inspiration from David Hume, David Lewis defended the view I am calling \emph{Neo-Humeanism}, i.e., that law statements merely provide a description of local matters of particular fact.\footnote{Whether Hume defended a view appropriately like Neo-Humeanism is beyond the scope of this essay; he probably did not.} In contrast, Anti-Humeans maintain that there are necessary connections in addition to the local matters of particular fact. In the following two subsections, I show that both Anti-Humean and Neo-Humean Neo-Russellians can explain $F$ while denying $C$. While I will remain neutral between Neo-Humeanism and Anti-Humeanism, I show that, for the Neo-Humean, the argument that $F$ cannot be explained without $C$ is a special case of an already well-known objection that Neo-Humeans have responded to at length.

\subsection{Anti-Humean Explanations}

Assuming an Anti-Humean account, the Neo-Russellian can endorse a logically stronger version of the TCP, in which all the particular facts must be consistent with various logical, metaphysical, and nomic principles because those principles constrain what is possible. Some Neo-Russellians -- such as Emily Adlam (\citeyear{Adlam:2022, Adlam:2023}) and Eddy Keming Chen and Sheldon Goldstein (\citeyear{ChenGoldstein:2022}) -- have already constructed Anti-Humean accounts of laws. As Chen and Goldstein (\citeyear[38]{ChenGoldstein:2022}) put their view, the laws ``govern the behavior of material objects by constraining the physical possibilities''. As Adlam (\citeyear[14]{Adlam:2022}) describes, constraints are not merely ``\emph{descriptions} of the way things'' happen -- as Neo-Humeans would insist -- but also dictate ``the way things \emph{must} happen''. Anti-Humean Neo-Russellianism is consistent with the view that whatever begins to exist is determined to do so, but also consistent with the view that whatever begins to exist is merely probabilified. A third possibility consistent with the view is that physical reality is constrained to evolve in one of several different ways but neither determined nor probabilified to do so.\footnote{For example, though General Relativity allows entities to randomly begin to exist at the ring singularity in a rotating black hole, General Relativity does not entail a probability distribution on what begins to exist.}

Adlam explicitly relates her view of laws to an older view of causation. According to the older view of causation, causes constrain what happens because causes have modal force. This is presumably the view taken by the HEPP advocate, since they think that without causation, nothing at all would constrain what can begin to exist. As an advocate of Neo-Russellianism-1 and -2, Adlam maintains that the laws do not include the kind of asymmetry that would allow an interpretation in terms of causation; moreover, the laws relate entire global states of the universe. She refers to her view as the `all-at-once' view of laws. Adlam notes that even without fundamental causation, there can still be relations with modal force:

\begin{quote}
    [...] if the `all-at-once' view of lawhood is accepted then there probably can’t be anything like causal structure at the most fundamental level, since events determined by all-at-once laws will  typically depend on one another in a reciprocal fashion so there will be no asymmetrical causal relations to be found. But the absence of specifically \emph{causal} structure need not defeat the modal structure approach to lawhood, because of course there could still be more general modal structure in the all-at-once setting (e.g. relations like metaphysical necessitation or ontological dependence), and therefore it should be a priority for proponents of modal structure to develop an account of these more general possibilities \cite[12]{Adlam:2022}.
\end{quote}

\noindent As Chen and Goldstein (\citeyear[39]{ChenGoldstein:2022}) write, ``Laws constrain the world by limiting the physical possibilities
and constraining the actual world to be one of them.'' For Anti-Humean Neo-Russellians, the Humean Mosaic is not only consistent with or entailed by a set of logical, metaphysical, and nomic principles, but -- in fact -- the Humean Mosaic \emph{must, of necessity}, satisfy a corresponding set of constraints. 

Let's consider examples of Preventive Facts explained by constraints. Non-causal factors preclude an infinitude of possible items from beginning to exist and can preclude them at some times but not at other times. For example, it may be that there are two (or more) entities that, as a matter of logical or metaphysical necessity, cannot exist together in one possible world; in that case, we'd have an explanation for at least some of the Preventive Facts. Suppose Susan essentially has the property of being the only child of Tom and Mary. (This is meant only as an illustrative example and not as a serious metaphysical proposal.) In that case, Carl, who, if he existed, would also essentially be the only child of Tom and Mary, cannot begin to exist in any world where Mary exists; moreover, while perhaps Tom's existence isn't precluded prior to Mary's birth, Tom's existence is precluded after Mary's birth. There may also be a posteriori necessities, discovered through the natural sciences, that explain why whole classes of entities cannot begin. For example, without the electron field, electrons couldn't begin to exist.

We can also consider some widely known cases where what entities can begin, as well as where or when they can begin, is constrained by the laws. For example, Newtonian dynamics allows for the construction of so-called \emph{space invaders} scenarios \cite[34-37]{Earman:1986}. In Newtonian dynamics, there is no upper limit to the speed with which objects can travel. An assembly of five or more particles, all subject to Newtonian gravitational forces, can fly out, reach infinity, and exit spacetime altogether in finite time \citep{SaariXia:1995}. Since Newtonian dynamics is time reversal invariant, five or more objects can enter spacetime, fly in from infinity, and meet up with us in finite time. Such entities begin for no reason, can begin in arbitrarily large numbers, and in vastly different kinds. There is a simple reason that our spacetime is not overrun by such invaders: our spacetime's intrinsic light cone structure provides an upper bound on particle velocities and hence prohibits their existence.\footnote{I've assumed that some version of Penrose's Cosmic Censorship Conjecture holds, roughly, the principle that naked singularities are not nomologically possible.} Hence, the intrinsic structure of spacetime itself explains a wide range of Preventive Facts.

An anonymous reviewer has objected that light cone structure might instead be understood as an implicitly causal notion, because light cone structure might be understood to tell us what can cause what. The Neo-Russellian can offer at least three replies. First, note that the way in which specific light cone structures rule out the possibility of space invader scenarios is not a causal notion. So, \emph{that} explanation is independent of whether the light cone structure tells us what can cause what, even if the light cone structure is implicitly causal. Second, Neo-Russellians should claim that the light cone structure tells us what can depend on what without being implicitly causal. That is, Neo-Russellians already adopt the view that there are non-causal determination relations; hence, the Neo-Russellian would say that, instead of telling us what can cause what, the light cone structure tells us what can determine what. Third, even if the light cone structure did tell us, in some sense, what can cause what, that is, even if the light cone structure serves to rule out causal relations that do not respect the light cone structure, it doesn't follow that the light cone structure rules in causal relations that do respect the light cone structure.

Some readers may still be unsatisfied. Perhaps they are tempted to object that, contrary to the notion that the intrinsic structure of spacetime provides a non-causal explanation of Preventive Facts, spacetime's intrinsic structure suggests the reality of microphysical causation. After all, physicists often refer to the light cone structure as the \emph{causal} structure and discuss the constraints imposed by such structure as \emph{causal} constraints; furthermore, in the early twentieth century, there was an attempt -- by Hans Reichenbach and others -- to use the light cone structure to explain time in terms of causation. Nonetheless, it's easy enough to see why Neo-Russellians shouldn't accept a causal gloss on light cone structure. For example, Carl Hoefer writes,

\begin{quote}
    As Russell (1913) argued, advanced physical theories — and GR [General Relativity] is a paramount example — do not actually use the word `cause' in any fundamental sense. [...] SR [Special Relativity] can, perhaps be given an axiomatic presentation in terms of possible causal connections, but GR certainly cannot, and Einstein's equations, while specifying the \emph{mathematical} relationship between matter ($T^{ab}$) and metric $g^{ab}$, do not (of course) use the word `cause', nor specify in which direction(s), if any, the cause–effect relationships go. Relativists use `causal' in discussions of determinism, horizons, singularities, and so forth, but the term is easily eliminable from those discussions without substantive loss \cite[702]{Hoefer:2010}.
\end{quote}

\noindent Hence, the determination relations found in GR are (arguably) precisely the sort already endorsed by many Neo-Russellians, i.e., symmetric relations of codetermination, and the use of the word `causal' can be interpreted as a historical or sociological quirk, easily erasable without consequence. Furthermore, as Hoefer goes on to note,

\begin{quote}
    [...] the equations of the theory equally support counterfactuals whose causal correlates we would prefer not to accept. For example (and this was one of Russell's main points), the fact that GR permits retrodiction as much as prediction means that `backtracking' counterfactuals are as commonplace as forward‐looking counterfactuals, but we want to embrace only forward causation. And when we set up typical what‐if‐things‐here‐now‐had‐been‐different counterfactuals using the initial‐value formulation of GR (the most natural way to proceed), the so‐called \emph{constraint equations} — which must be satisfied if the theory is to be used at all — entail that changes in spacetime and/or matter \emph{here‐now} not only entail differences \emph{here‐later}, but also differences \emph{there‐now}, where `there‐now' may be a region spacelike separated from \emph{here‐now}. But we do not believe, and will not accept, that things here causally influence those distant states of affairs \cite[703]{Hoefer:2010}.
\end{quote}

\noindent Consequently, far from saving our ordinary causal notions, the intrinsic structure of spacetime imposes new challenges friendly to a Neo-Russellian analysis. The so-called `causal structure of spacetime' is a misnomer, at least insofar as that name suggests anything like efficient causation, and we end up with a new collection of determination relations undreamt of from the metaphysician's armchair. And this new collection of apparently non-causal determination relations helps to explain a variety of Preventive Facts in non-causal terms.

It may also be that if a sufficiently wide range of Preventive Facts have non-causal explanations, there need be no independent explanation for the Selective Facts. However, our best scientific theories do provide us with non-causal explanations of various Selective Facts. As I've said, whether or not particles begin to exist -- and so whether or not the objects supervening on field configurations begin to exist -- depends on how their fields evolve. In turn, the dynamical laws satisfied by the fields explain which items are either prevented from or selected for existence. In general, which evolutions of a field are possible is subject to various kinematic constraints. By way of analogy, consider the Seven Bridges of Königsberg. In the eighteenth century, Königsberg was divided into four areas connected by seven bridges. The problem is then to determine whether there exists a way of traversing all seven bridges without doubling back on oneself. As Euler proved, there is no trajectory that does so. The fact that there is no trajectory that does so is widely understood to have a non-causal graph theoretic explanation \citep{Reutlinger:2017}. Hence, independent of whatever may be true about causation, the possible trajectories through eighteenth century Königsberg were limited to trajectories that doubled back on themselves. Likewise, again independent of whatever may be true about causation, there are kinematic constraints on the evolution of various fields that either prevent various entities from coming into existence or that select various other entities for existence.

Consider the Glashow-Iliopoulos-Maiani mechanism. (For an introduction to this topic, see \citep[326-329]{Griffiths:2014}.) Before discovering the charm quark, physicists expected neutral kaons to decay into muon pairs at a detectable rate. However, such decays evaded detection; the failure to detect such decays could be explained by the existence of the charm quark. If the up and charm quarks had identical masses, the amplitudes for the two decay channels would destructively cancel; neutral kaons would never decay into muon pairs. The masses are not exactly equal, so kaons do decay into muon pairs, but at a much lower rate than was detectable. Hence, quantum field theory offers an explanation for various Preventive and Selective Facts.

As another example, consider the formation of a white dwarf star.\footnote{This example has previously been discussed as an example of a non-causal explanation in \citep{Colyvan:1999, Strevens:2011, Pexton:2016, Pexton:2017, FrenchSaatsi:2018}. The example originates in a comment that Peter Railton apparently made to David Lewis (\citeyear{Lewis:1987}). Lewis argued that although white dwarf formation does not have a cause, white dwarf formation does have a causal explanation. Lewis's account of `causal explanation' entails that any explanation \emph{ruling out} causation is a causal explanation. For my purposes in this paper, white dwarf formation has a non-causal explanation because (i) the explanation rules out causation and (ii) non-causal factors play the primary explanatory role.} Stars that fall into a specific range of masses, having burnt all of their fuel, begin to collapse upon themselves. Due to the fact that the electron wavefunction is anti-symmetric, no two electrons can exist in the same quantum state. This provides a fundamental kinematic constraint on the evolution of the electron field. And given that kinematic constraint, stars in a specific range of masses halt in their collapse; the result is a white dwarf star. In a less than careful description of the formation of a white dwarf star, we might say that the stellar material collapsed due to gravitational forces, and that the collapse was halted by the pressure generated due to the fact that no two electrons can occupy the same state. Described that way, the formation of a white dwarf star appears to have a causal explanation. However, a more careful description would point out that when we say that objects are moving under the influence of gravity, we really mean that the objects are following the motions natural for them given the inertial structure of spacetime; when we say that their motion is arrested by the pressure due to the fact that no two electrons can occupy the same state, we really mean that their motion is subject to a kinematic constraint. In other words, the formation of a white dwarf star has a non-causal explanation. Hence, quantum mechanics and General Relativity, together, provide a non-causal explanation for one of the Selective Facts, i.e., that white dwarf stars form under a specific set of conditions, and for specific Preventive Facts, e.g., that, under those conditions, nothing else results.

Steven French and Juha Saatsi (\citeyear[192-194]{FrenchSaatsi:2018}) have argued that both ionic and covalent bonding -- and so the formation of molecules -- should be understood to have a non-causal explanation in terms of a similar kinematic constraint on multi-particle wavefunctions. In that case, provided the appropriate arrow of time is available, failures of $C$ are not limited to exotic situations, such as the formation of white dwarf stars or the decay of subatomic particles, but are ubiquitous; failures of $C$ appear whenever a new molecule appears. In that case, instances where objects begin to exist for non-causal reasons \emph{vastly} outnumber cases where they begin for causal reasons. Again, we have an explanation both for some Selective Facts -- that specific molecules begin in specific circumstances -- and for some Preventive Facts -- that other physical items do not begin in those circumstances.

As yet another example, consider the close relationship between symmetries and conserved quantities that follows from Noether's Theorem. For example, the fact that the electron field locally satisfies phase symmetry entails the conservation of electric charge. In turn, the conservation of electric charge entails that there is no state of affairs where (for example) zero net charge enters and non-zero net charge exits. Hence, we once more have a non-causal explanation of various Preventive and Selective Facts.

Furthermore, there are statements plausibly interpreted as expressing laws, such as the Einstein Field Equation, but which are not naturally interpreted as stating laws about how the world evolves from, e.g., one Cauchy surface onward.\footnote{There are three interrelated issues. First, while there is some sense in which General Relativity has an initial value formulation, i.e., the Hamiltonian formulation, the Hamiltonian formulation places controversial restrictions on the theory's solution space by, e.g., restricting the theory to globally hyperbolic solutions. Second, as Chen and Goldstein (\citeyear[46]{ChenGoldstein:2022}) note, many globally hyperbolic spacetimes do not have a unique slicing into Cauchy surfaces. Third, while the Hamiltonian formulation may be analogous to the initial value formulation of other physical theories, the Hamiltonian formulation is notoriously difficult to interpret in any straightforwardly or intuitively temporal terms, c.f. \citep{Earman:2002}. In General Relativity's standard formulation, the Einstein Field Equations are solved by specifying (for example) the global matter-energy distribution, over the entire spacetime, which then fixes the metric tensor for the entire spacetime.} Such laws constrain which global states of the world are possible but resist a straightforward interpretation in which events at one time constrain those at later times by causing them. Because they non-causally constrain the global state of the world, sans $C$, such laws can prevent some range of items from coming into existence and constrain which items could be selected for existence.

Consequently, even without $C$, there would still be  explanations Anti-Humeans would accept for $F$. In the next subsection, I show that, given a Neo-Humean account of laws, there would also be explanations Neo-Humeans would accept for $F$. For Neo-Humean Neo-Russellans, the claim that, without $C$, there would be no good explanation for $F$ turns out to be a special case of an objection Neo-Humeans have already responded to at length.

\subsection{Neo-Humean Explanations}

In order to explicate Neo-Humeanism, let's begin with natural qualities. Natural qualities, such as having electric charge, can be distinguished from non-natural qualities, such as the disjunctive quality of having electric charge or being the Eiffel Tower.\footnote{This example is from \cite[27-28]{ChenGoldstein:2022}.} Some qualities are more natural than others. Neo-Humeans maintain that there is only the Humean Mosaic, that is, spacetime and the distribution of perfectly natural qualities across spacetime. Everything else -- laws, chance, causation -- supervenes on that distribution. Since each instance of a perfectly natural quality is independent of every other, there are no necessary connections between distinct existences. Instead, ``all there is to the world is
a vast mosaic of local matters of particular fact, just one little thing and
then another'' \citep{Lewis:1986b}. On the most popular Neo-Humean account of laws -- the \emph{Best System Account} -- laws are the axioms of the Best System, that is, the deductive system that both describes the Humean Mosaic and makes the best trade-off truth will allow between simplicity, informativeness, and fit \citep[478, 480]{Lewis:1994}. For Neo-Humeans, a large spatiotemporal region $E_1$ -- such as an instant of time, a Cauchy surface, or whatever -- determines another $E_2$ just in case, given $E_1$ and the laws of nature, $E_2$ follows as a deductive consequence \citep{Lewis:1973, Lewis:1981, Loewer:2008}. Furthermore, as with the Anti-Humeans, Neo-Humeans do not need to talk about one time, Cauchy surface, or whatever, determining any other; Neo-Humeans can endorse all-at-once laws. Furthermore, Neo-Humean analogues of the various explanations offered in the previous section -- for the prevention of space invaders, the selection of white dwarf stars or molecules, and so on -- can obviously be offered.

Suppose someone asks why, at the closest possible world where there is no microphysical causation, entities do not begin to exist in arbitrary numbers, of arbitrary kinds, or in arbitrary spatiotemporal locations. The world closest by is the one where the distribution of local matters of particular fact is closest to the actual world's distribution. While those worlds may include violations of the laws -- that is, miracles -- the worlds that are closest by are those with the fewest and smallest such violations -- that is, the fewest and smallest miracles. Consequently, even without microphysical causation, arbitrary numbers and kinds of things cannot begin to exist at arbitrary spatiotemporal locations because their doing so would represent a gross violation of natural law.

The HEPP proponent might reply as follows. If $C$ were false, then, with respect to any given Cauchy surface (or whatever), the vast majority of future continuations of the Humean Mosaic are disorganized, with completely arbitrary entities beginning to exist. Hence, without $C$, the most probable continuation would involve arbitrary entities beginning to exist. Therefore, the argument might go, the Best System Account doesn't adequately explain the Preventive and Selective Facts after all. Call this the Chaotic Continuation Objection (CCO).

The CCO ignores how friends of the Best System Account understand chances. For friends of the Best System Account, chances are grounded in the distribution of natural properties over the entire Humean Mosaic. Rational credences are constrained by chance via the Principal Principle; since there is little (or no) chance for arbitrary entities, in arbitrary numbers, to begin for no reason at all over the next second, our credence in such a possibility should be very low. Moreover, friends of the Best System Account would argue that the CCO has confused metaphysical independence with probabilistic independence. While Neo-Humeans claim that each instance of a perfectly natural property is metaphysically independent, they deny that the instances are probabilistically independent \cite[136]{Loewer:2023}. Hence, since we've never seen a violation of natural law, we should have a high credence that we will not be deluged by completely arbitrary entities over the next second.

Perhaps the HEPP proponent can claim that, had $C$ been false, typical Humean Mosaics are such that no good trade-off between simplicity, informativeness, and fit is possible. If so, perhaps, without $C$, we shouldn't expect order over the next second. But in virtue of what can this claim be made? In statistical mechanics, the claim that some state is atypical is usually made with respect to a measure; since we have a theory of the underlying dynamics, some measure (e.g., the Liouville measure for Newtonian dynamics) is picked out as a particularly natural choice. With respect to the collection of possible Humean Mosaics, we lack a theory to inform our choice of measure \cite[135]{Loewer:2023}. Without a theory to inform our choice of measure, Best System Account proponents would say that there are no grounds for saying that typical Humean Mosaics have any specific, contingent feature.

Furthermore, it's unclear how $C$ secures an advantage in explaining the Preventive and Selective Facts over the account already provided by the Best System Account without $C$. In comparing the Best System and standard Anti-Humean accounts, Loewer (\citeyear[138]{Loewer:2023}) argues that neither provides an advantage in ruling out disorganized Humean Mosaics. If Lewis were right that there are no necessitation relations, every distribution of natural properties over the Humean Mosaic would have been metaphysically possible. But if a standard Anti-Humean account had been correct, the space of metaphysically possible worlds would be populated by every possible necessitation relation; hence, for every possible world that obtains under the former analysis, a corresponding and just as disorganized world obtains under the latter.

The same holds for causation. Even if entities could only be selected for or prevented from existence by causes, we'd have no assurance our world would be ordered. Causes and their effects are linked by a set of law-like connections. For example, that specific interventions on a pile of wood result in a house obtains only because specific law-like connections obtain. With other law-like connections, attempting the same (or similar) intervention on a pile of wood yields a different result. Hence, even with a restriction to worlds where $C$ holds, we have little reason for thinking the ordered intervention/outcome combinations will dominate over the unordered intervention/outcome combinations.

Nonetheless, one could object that a mere summary of the distribution of perfectly natural qualities over the Humean Mosaic fails to explain why there is such a distribution in the first place. Hence, they would claim the Best System Account is incapable of providing a good explanation for why things do not begin at arbitrary places, times, or in arbitrary number or kind. Alternatively, one could object that without necessary connections between distinct existences, given the distribution of perfectly natural qualities over some region of the Humean Mosaic, there is no good reason to expect any specific distribution of perfectly natural qualities over any other region of the Humean Mosaic. In their view, if there are no necessary connections between distinct existences, we would be left without an otherwise available solution to the problem of induction. In that case, they might claim, there can be no satisfactory explanation for why entities do not begin at arbitrary times or locations or in arbitrary number or kind.

Friends of the Best System Account find these objections question begging. Metaphysical explanation must bottom out somewhere. Perhaps metaphysical explanation bottoms out in the distribution of perfectly natural qualities on the Humean Mosaic. Here, I don't take a stand on whether good explanations of $F$ can be offered by the Anti-Humean. Instead, it suffices to note that Neo-Russellians can rest upon another shore; Neo-Russellians can endorse necessary connections between distinct existences by either modifying the Best System Account to include necessary connections (as in Loewer's (\citeyear{Loewer:2021}) Package Deal Account) or else by endorsing an Anti-Humean account of laws. 

\section{Conclusion}

According to the HEPP, if anything could begin without a cause, so that the Causal Principle is false, there'd be no good explanation for why entities do not begin at arbitrary times, in arbitrary spatial locations, in arbitrary number, or of arbitrary kind. While the HEPP can be used to construct an objection to Neo-Russellianism, Neo-Russellians have good reason to reject the HEPP. For Neo-Russellians, the Causal Principle is either false or vacuously true. Either way, in the nearest worlds where the Causal Principle is non-vacuously false, what begins to exist is logically, metaphysically, and nomically constrained. In turn, those constraints explain why entities do not begin at arbitrary times, in arbitrary spatial locations, in arbitrary number, or of arbitrary kind. Neo-Russellians can make use of widely recognized non-causal explanations of various entities (e.g., subatomic particles, white dwarf stars, molecules, etc) beginning to exist that are easily interpreted in terms of either an Anti-Humean or Neo-Humean account of laws. Lastly, objections to Neo-Humean Neo-Russellianism based on the HEPP turn out to be a special case of a standard objection to Neo-Humeanism and for which there already exist standard responses.

\bibliographystyle{apa}
\bibliography{references.bib}

\end{document}